# Analysing the Public Discourse around OpenAI's Text-To-Video Model 'Sora' using Topic Modeling.  (1472 words)

## 1. Introduction:

The rapid advancements in generative artificial intelligence (gen AI) models have sparked widespread public discourse and debate, with the latest development being OpenAI's text-to-video model, Sora. Announced on February 15, 2024, it instantly caught the public's attention by demonstrating the ability to generate dynamic and realistic video clips from text prompts, similar to how OpenAI's DALL-E generates images from text. While Sora is still in a pre-release phase, its potential to revolutionize content creation and disrupt various industries be it media, entertainment, or advertising, has already ignited discussions across online communities.

Subreddits such as r/OpenAI, r/technology and r/ChatGPT have emerged as epicentres for technology enthusiasts and critics to openly discuss and share narratives about the latest advancements in AI technologies. Previous studies have explored public perceptions of large language models like ChatGPT and image generators such as DALL-E through analysing online forums. For instance, Talafidaryani and Mora (2024) employed topic modeling techniques on Reddit data to uncover dominant themes surrounding ChatGPT, including its capabilities, limitations, and ethical considerations. Similarly, Zhou and Nabus (2023) investigated discussions on DALL-E, revealing discourse on creative applications, risks of misuse, and comparisons to human artists. However, due to Sora's relatively recent emergence, there is still a lack of research on the narratives and themes emerging from Reddit conversations about this novel technology.

By conducting topic modeling analysis on a large corpus of Reddit comments, the study aims to feel that gap and uncover the main topics and themes users are discussing about Sora. These narratives can provide valuable insights into public perceptions, areas of excitement, as well as societal and ethical concerns surrounding around the advent of new generative AI technologies.

## 2. Research Question:

What are the dominant themes and topics discussed by users in the Reddit communities regarding OpenAI's latest text-to-video model 'Sora'?

## 3. Method

### 3.1. Data:

For this study, Reddit comments discussing Sora were extracted from five subreddits: r/OpenAI, r/technology, r/singularity, r/vfx and r/ChatGPT. These subreddits were chosen because they are active online communities discussing the latest advancements in generative AI technologies. The data was collected over a 2-month period, from February 1, 2024, to April 1, 2024, to capture the initial public discourse surrounding the announcement and release of Sora.

 The data collection process involved two steps. Firstly, Reddit posts containing the case-insensitive keywords "Sora" and "OpenAI" in their title or body text were manually identified based on the "Relevance" filter on the Reddit website. Two posts were then selected from each subreddit with the highest number of comments, with the minimum comment count being 500. Next, using the Python Reddit API Wrapper (PRAW) library, the top 200 comments in English language were extracted from each of the identified posts, based on their score or number of upvotes. This resulted in a total corpus of 2,000 comments across the ten selected posts.

Each data point in the dataset included the comment text, its timestamp, number of upvotes (score) and metadata such as the post title, author's unique id, and its subreddit source.

**3.2. Analysis:**

The collected Reddit comments underwent a series of preprocessing steps to prepare the data for analysis. Firstly, the comments were converted to lowercase, and non-alphanumeric characters, URLs, and specific words like 'http', 'www', and 'com' were removed using regular expressions. The text was then tokenized, splitting it into individual words or tokens. Stop words, such as 'the', 'a', and 'and', were then removed from the tokenized text, as they do not contribute significantly to the meaning of the comments. Next, lemmatization was performed on the remaining tokens using the WordNetLemmatizer from the NLTK library. After refining the dataset, 1827 comments were selected and converted to a Pandas dataframe.

After preprocessing, feature extraction was carried out on the corpus of comments and specifically, the Term Frequency-Inverse Document Frequency (TF-IDF) technique was employed using the gensim library. TF-IDF evaluates the importance of a word within a document and assigns higher weights to words that are more relevant and informative. Topic modeling was then implemented using the Latent Dirichlet Allocation (LDA) algorithm (Blei et al., 2003) and coherence scores were computed for different values to determine the optimal number of topics (k) for the LDA model. Four was the optimal number to get coherent topics and ten words were displayed per topic. To make the topics more interpretable, titles were assigned based on the most representative words in each topic.

| **Comment Text** | **Date** | **Score** | **Subreddit** | **Pre-processed Text** |
|---|---|---|---|---|
| Quality is insane. Stock footage companies will now be on life support. | 2024-02-15 19:02:03 | 576 | OpenAI | quality insane stock footage company life support |
| The planet is dying from climate change. The best big tech can do is give teenagers feature films about their waifus. | 2024-02-16 19:33:00 | 473 | technology | planet dying climate change best big tech give teenager feature film waifus |
| Man, wtf, the quality is insane. How is this technology not supposed to make millions of people jobless? | 2024-02-15 18:28:45 | 227 | OpenAI | man wtf quality insane technology supposed make million people jobless |

**Table 1. Subset of Data after Cleaning and Preprocessing**

To visualize and interpret these topics, word clouds were generated to display the most frequent words associated with each topic. Additionally, bar charts were created to illustrate the relative importance of topic keywords plotted against word count and understand the dominant themes with their respective weights in the conversations. Further analysis was conducted using dimensionality reduction techniques like t-SNE to cluster and visualize the topic distributions across the comments, along with pyLDAvis for interactive visualization of the LDA topic modeling results.

**4. Results:**

The topic modeling analysis identified four key topics, which were further interpreted and labelled based on the most prevalent words associated with each topic as shown in Table 2. Topic 1, titled "AI Impact and Trends in Sora Discussions", focused on the broader impact of AI technologies and their evolving trends with keywords like "ai", "human", "future", and "job" indicating conversations about the potential effects of Sora on employment and industries. Topic 2, "Public Opinion and Concerns about Sora", captured the discussions around public perceptions, sentiments, and ethical considerations regarding the new model, as evidenced by words such as "people", "think", "artist", and "video".

Topic 3, "Artistic Expression and Video Creation with Sora", highlighted the excitement and potential of using Sora for creative applications, with keywords like "art", "video", "creative", and "artist". Finally, Topic 4, "Sora's Applications in Media and Entertainment", reflected discussions on the possible use cases of Sora in various media and entertainment industries, as indicated by words like "model", "movie", "real", and "work".

| Topic | Topic Name | Words |
| --- | --- | --- |
| Topic 1 | AI Impact and Trends in Sora Discussions | ai, people, going, year, get, like, think, thing, job, artist |
| Topic 2 | Public Opinion and Concerns about Sora | people, ai, think, like, year, need, want, video, even, one |
| Topic 3 | Artistic Expression and Video Creation with Sora | ai, like, art, video, people, get, make, artist, time, going |
| Topic 4 | Sora's Applications in Media and Entertainment | ai, people, like, make, video, model, one, job, movie, work |

**Table 2. Topics Determined by LDA along with Most Prevalent Words**

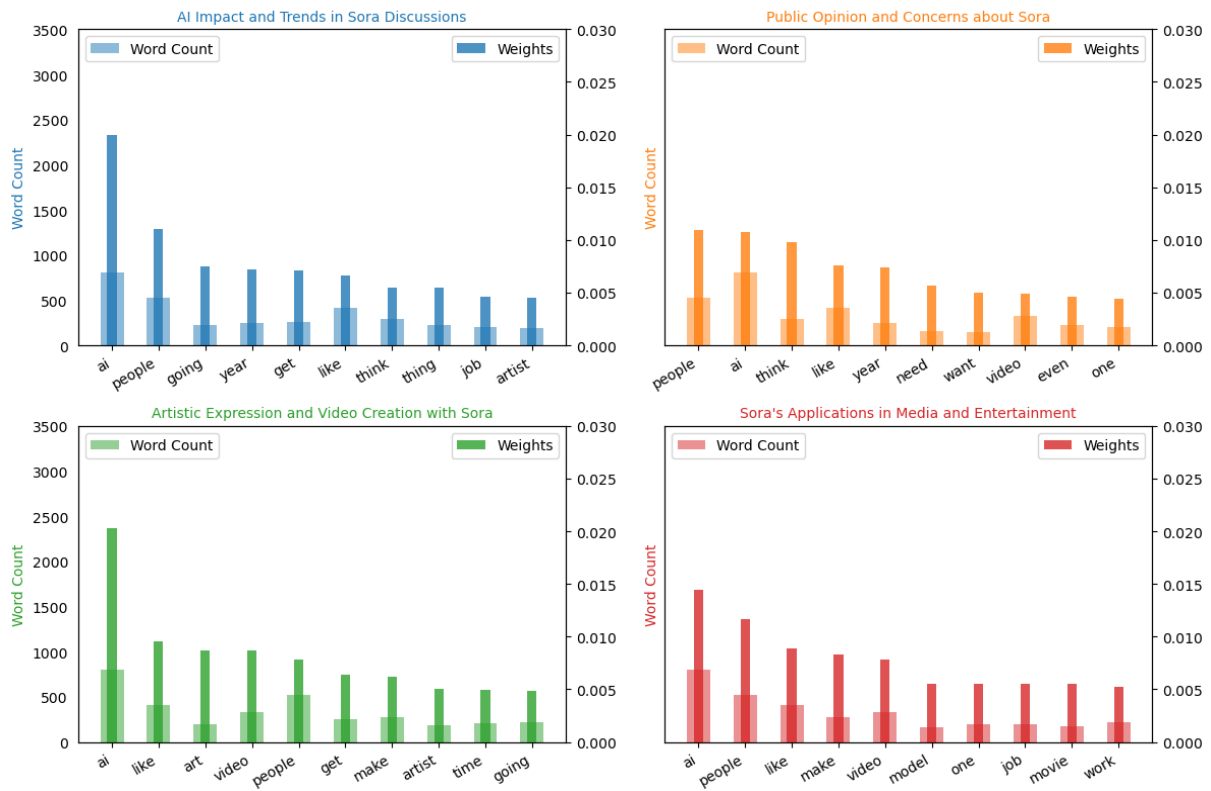

**Figure 1. Bar Charts showing Word Count and Importance of Topic Keywords**

The bar charts in Figure 1 provide insights into the word count (frequency) and importance (weights) of the keywords associated with each identified topic. Across the four topics, certain words such as "ai", "people", "job", "artist" stand out as highly frequent and influential within their respective topics. The word clouds in Figure 2 offer a visual representation of the most prominent keywords associated with each topic.

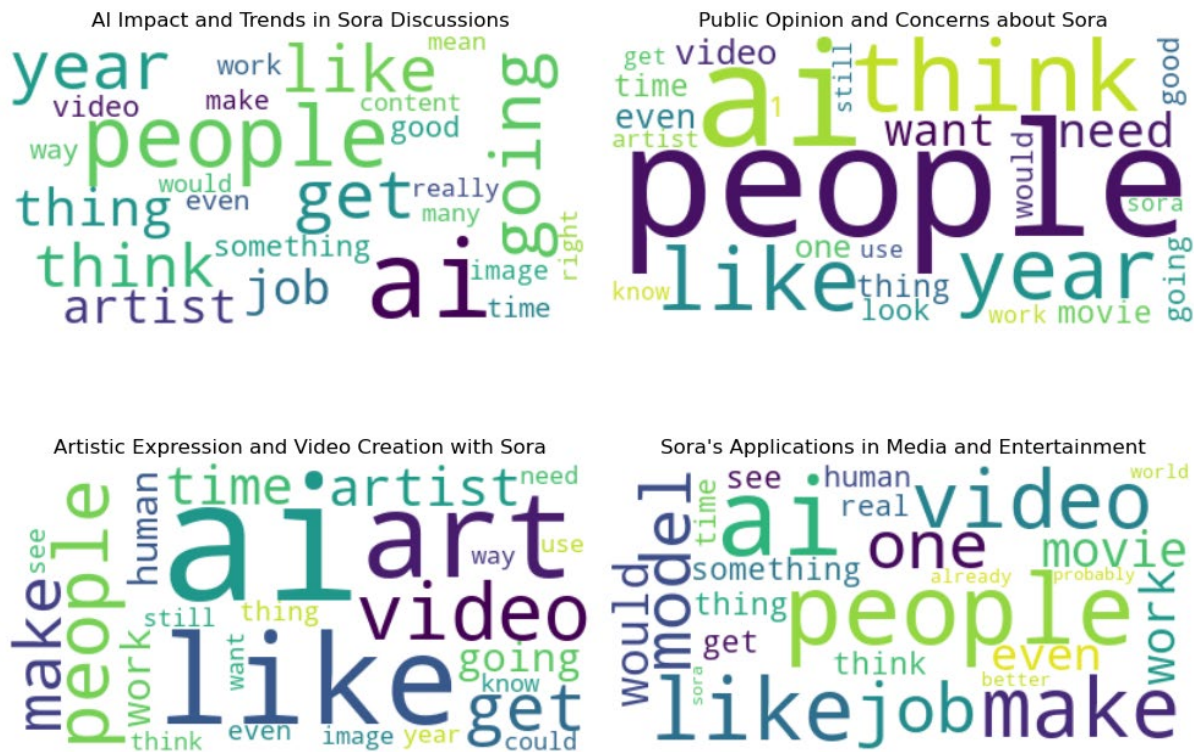

**Figure 2. Word Clouds Visualizing Topic Keywords**

Figure 3 provides an interactive visualization of the LDA topic modeling results using the pyLDAvis library. The intertopic distance map shows the similarity or dissimilarity between the identified topics. It can be inferred that Topics 1 and 3 are overlapping, indicating that user discussions converge on themes such as the impact of AI and its creative applications. The bar chart in Figure 3 displays the top 30 most relevant terms for each topic, and interestingly, Topic 2 had the highest average relevance score (0.0203) across its top 30 terms. This suggests that the keywords used in discussing concerns about Sora are more likely to be occurring frequently throughout the topic.

The t-SNE (t-Distributed Stochastic Neighbour Embedding) clustering in Figure 4 depicts the distribution of comments across the identified LDA topics. Each point in the plot represents a comment, and different colours indicate the topic cluster each comment belongs to, based on its dominant topic. The clustering patterns observed suggest that Topics 1 and 3 (blue and green) are more closely related whereas Topics 2 and 4 (orange and red) appear more distinct and separable indicating divergent narratives within the Reddit discussions surrounding Sora.

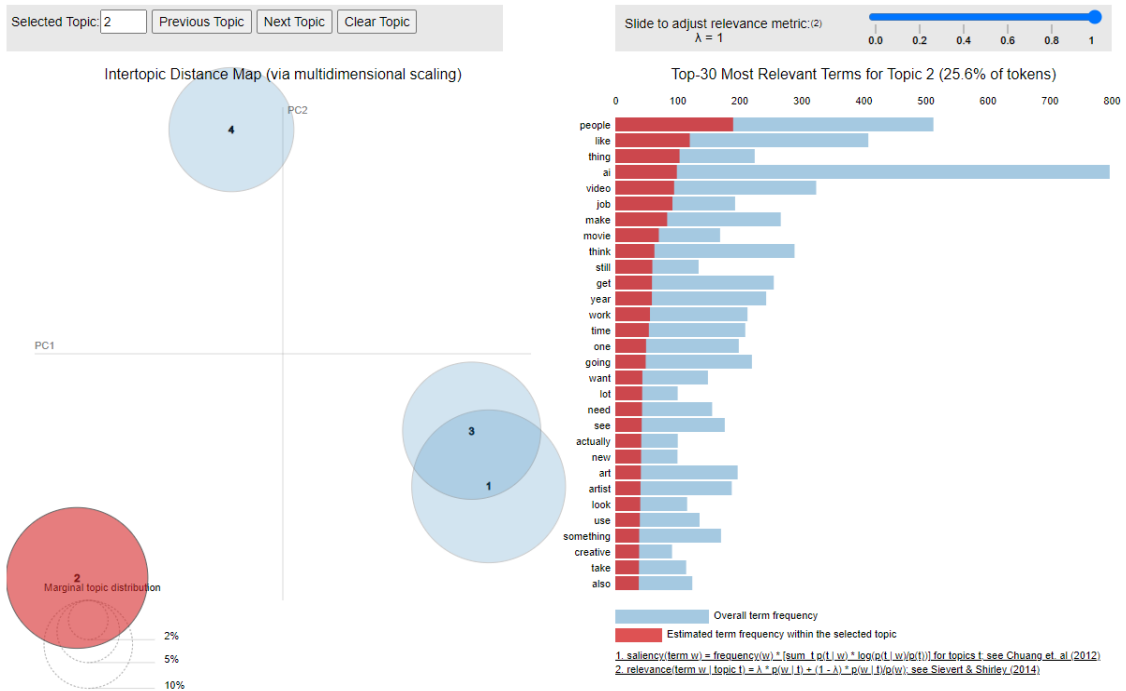

**Figure 3. LDA Topic Visualization and Term Relevance using pyLDAvis**

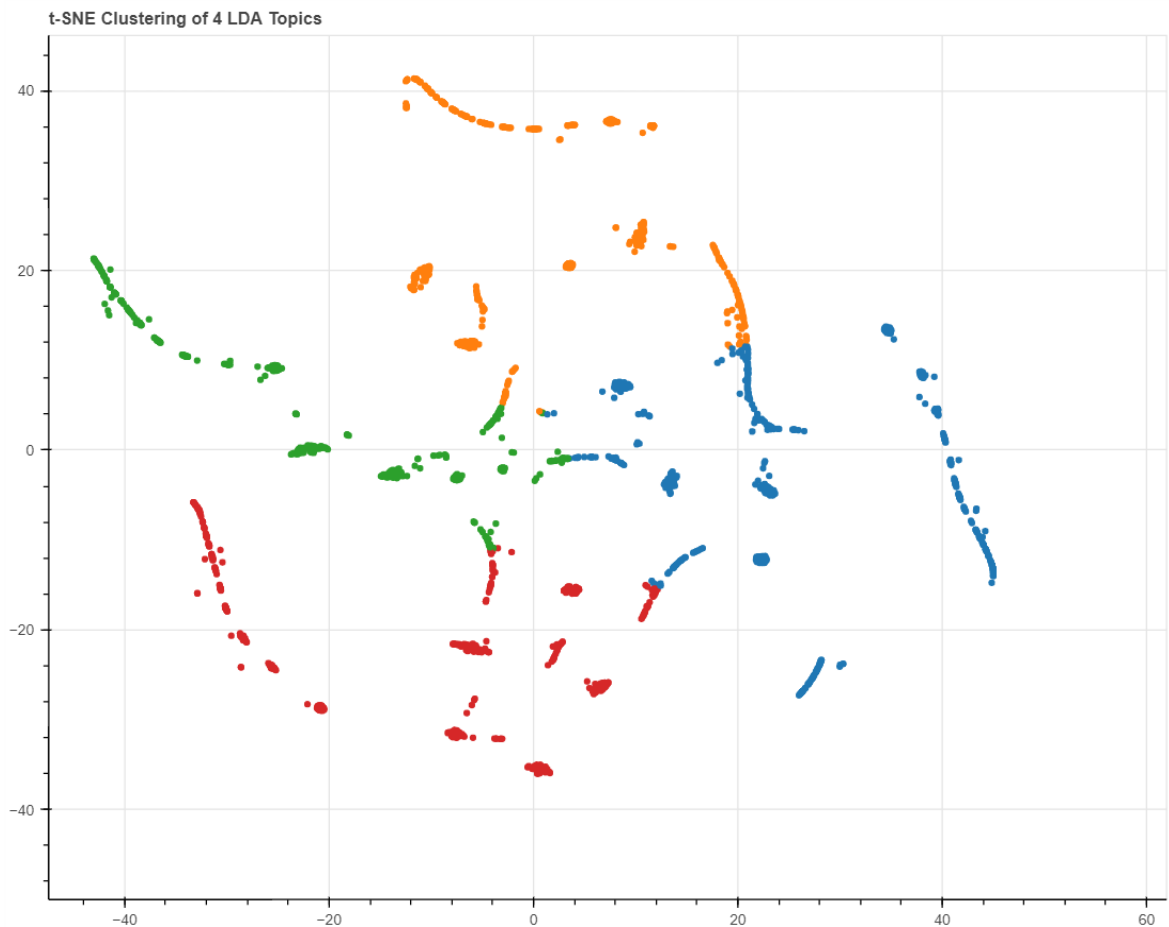

**Figure 4. t-SNE Clustering of LDA Topics**

## 5. Conclusion and Limitations:

The topic modeling analysis performed on the Reddit comments aimed to uncover the dominant themes and topics discussed by users in online communities regarding OpenAI's text-to-video model Sora. The results, as presented in the previous section, successfully identified four key topics that directly address the research question. These topics capture discussions around the broader impact of AI technologies like Sora, public opinions and ethical concerns, creative applications and artistic expression enabled by the model, and its potential use cases in various industries, particularly media and entertainment. Visualizations such as bar charts, word clouds, and t-SNE clustering further helped understand the distribution of topics and the relevance of keywords within each theme.

The study's primary limitation lies in its reliance solely on textual data from Reddit comments, overlooking other social media platforms such as Twitter, Instagram or Facebook. Moreover, the interpretations of topic labels and the assignment of representative words were subjective processes, potentially influenced by individual biases. Also, the study was restricted to a subset of five subreddits (r/OpenAI, r/technology, r/singularity, r/vfx, and r/ChatGPT), a fixed number of top-voted comments and restricting to a specific two-month timeframe. This may not capture the full scope of narratives expressed across other subreddits, outside the chosen time period or from less popular comments. Despite these limitations, the study provides a framework for future research to build upon in capturing public perceptions surrounding Sora and incorporating sentiment analysis to further gain insights into the emotional tones associated with the identified topics.